\documentclass[conference]{IEEEtran}
\IEEEoverridecommandlockouts
\usepackage{cite}
\usepackage{algorithmic}
\usepackage{textcomp}
\usepackage{xcolor}
\def\BibTeX{{\rm B\kern-.05em{\sc i\kern-.025em b}\kern-.08em
    T\kern-.1667em\lower.7ex\hbox{E}\kern-.125emX}}
\usepackage{amsmath}
\usepackage{physics}
\usepackage{amsfonts}
\usepackage{amssymb} 
\usepackage{braket}
\usepackage{xfrac} 
\usepackage{leftidx} 
\usepackage{tikz} 
\usepackage{subfiles}
\usepackage[colorlinks=true, allcolors=blue]{hyperref}
\usepackage{graphicx}
\usepackage{subfig} 
\usepackage{svg} 
\usepackage{comment}

\newcommand{\ceil}[1]{\lceil {#1} \rceil}

\begin{document}

\title{An efficient algebraic representation for graph states for measurement-based quantum computing
}

\author{\IEEEauthorblockN{Sebastiano Corli}
\IEEEauthorblockA{\textit{Dipartimento di Fisica}}
\textit{Politecnico di Milano}\\
Milan, Italy \\
\IEEEauthorblockA{\textit{Istituto di Fotonica e Nanotecnologia} \\
\textit{Consiglio Nazionale delle Ricerche}\\
Milan, Italy \\
sebastiano.corli@polimi.it}
\and
\IEEEauthorblockN{Enrico Prati}
\IEEEauthorblockA{\textit{Dipartimento di Fisica}}
\textit{Università degli Studi di Milano}\\
Milan, Italy \\
\IEEEauthorblockA{\textit{Istituto di Fotonica e Nanotecnologia} \\
\textit{Consiglio Nazionale delle Ricerche}\\
Milan, Italy \\
enrico.prati@unimi.it}
}

\maketitle

\begin{abstract}
Graph states are the main computational building blocks of measurement-based computation and a useful tool for error correction in the gate model architecture. The graph states form a class of quantum states which are eigenvectors for the abelian group of stabilizer operators. They own topological properties, arising from their graph structure, including the presence of highly connected nodes, called hubs. Starting from hub nodes, we demonstrate how to efficiently express a graph state through the generators of the stabilizer group. We provide examples by expressing the ring and the star topology, for which the number of stabilizers reduces from $n$ to $\ceil{\frac{n}{2}}$, and from $n$ to $1$, respectively.
We demonstrate that the graph states can be generated by a subgroup of the stabilizer group. Therefore, we provide an algebraic framework to manipulate the graph states with a reduced number of stabilizers.
\end{abstract}


\section{Introduction}
Measurement-based quantum computing (MBQC) represents a powerful computational architecture based on the virtualization of quantum circuits, specially suitable  for \cite{briegel2009measurement, adcock2019programmable, scott2022timing} -- but not limited to \cite{wang2020integrated, albarran2018one,lanyon2013measurement} -  the photonic technology. Many platoforms for quantum computing (trapped ions~\cite{blatt2012quantum, de2022temperature}, ultracold atoms~\cite{wigley2016fast}, superconducting qubits~\cite{krantz2019quantum}, spin qubits in silicon~\cite{rotta2017quantum}) require cryogenic temperatures to be shielded against noise and decoherence, however photonic chip work at room temperature\cite{aharonovich2016solid,grosso2017tunable, dietrich2020solid}, making photonic platform potentially portable.

While the gate-based model seeks to reproduce coherent operations on a quantum hardware, the measurement-based paradigm virtualizes the circuit over a number of physical operations conditioned by measurements carried on single qubits at each step of the circuit. Such process allows to avoid two-qubits gates, making the process easier than implementing coherent operations~\cite{zwerger2014hybrid, menicucci2014fault}. Therefore, instead of manipulating coherent states, the MBQC model aims to manipulate a set of entangled source states, known as graph states. The graph states and the stabilizer group provide a powerful computational tool to involve the entanglement between qubits in the measurement-based computation~\cite{adcock2019programmable, raussendorf2003measurement ,fujii2015quantum,briegel2009measurement, hein2004multiparty, adcock2020mapping, nikahd2015one, morimae2014acausal, guhne2005bell}, \cite[pag.4]{hein2006entanglement} and to encode error correction codes~\cite{adcock2019programmable,hein2004multiparty,briegel2009measurement,raussendorf2003measurement,fujii2015quantum,egan2020fault, schlingemann2001quantum, chiaverini2004realization,devitt2013quantum, gottesman2002introduction,roffe2019quantum,hsieh2007general,guenda2018constructions,nautrup2019optimizing,hein2006entanglement,gottesman1997stabilizer}. As for gate model and adiabatic architectures, efficient preparation of quantum states \cite{maronese2022quantum,maronese2022quantum2} is of major importance in order to reduce the classical to quantum conversion overhead.
Here, we show an efficient expression of graph states through the generators of the stabilizer group. The resulting algebraic framework allows to manipulate the graph states with a reduced number of stabilizers. In the final part we apply it to the cases of the ring and the star topology. 

\section{Computing by graph states}

\begin{figure}
\subfloat[\label{subfig:GHZ}]{{\includegraphics[width=6cm]{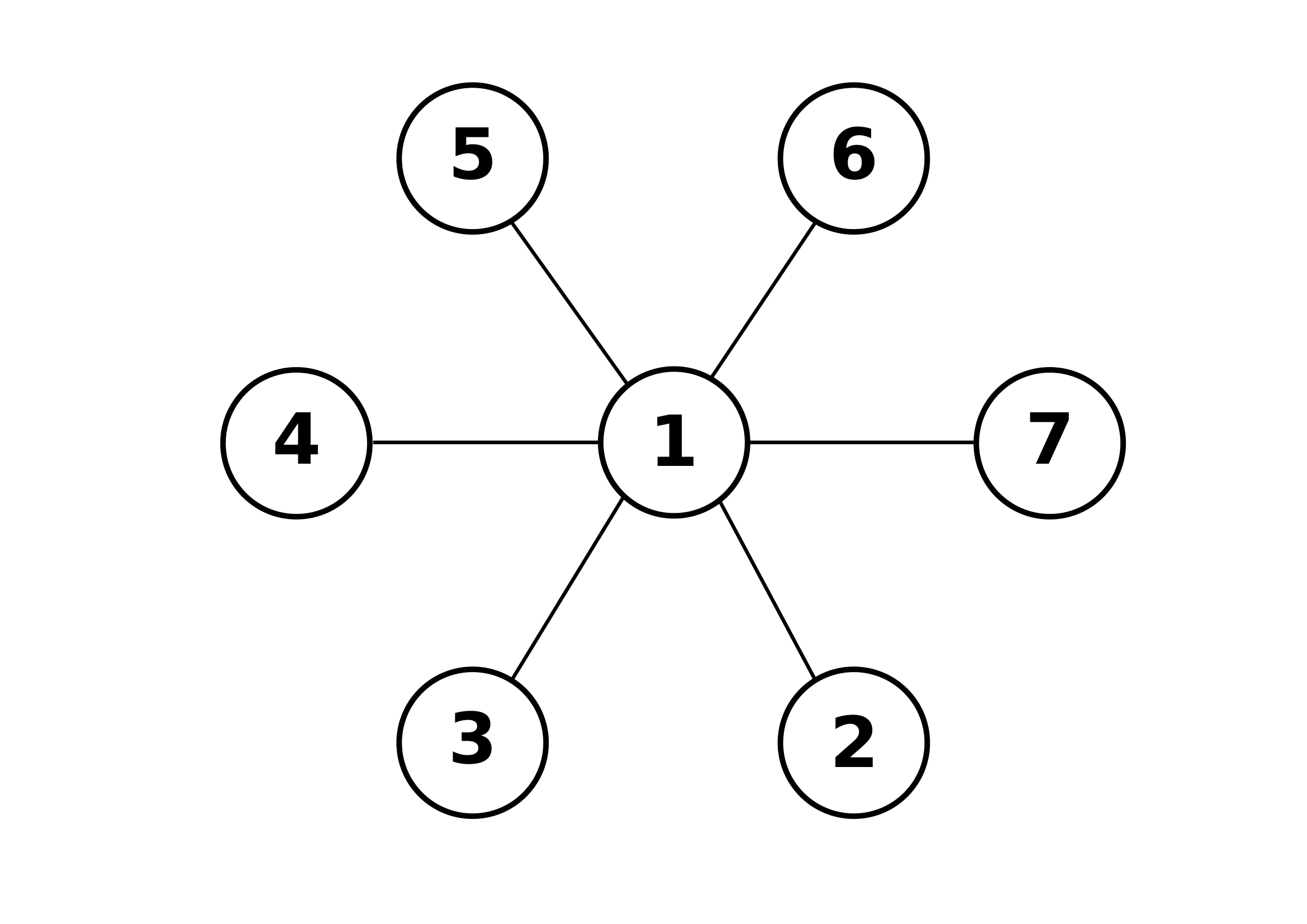} }}
\quad
\subfloat[\label{subfig:CHZregister}]{{\includegraphics[width=5.53cm]{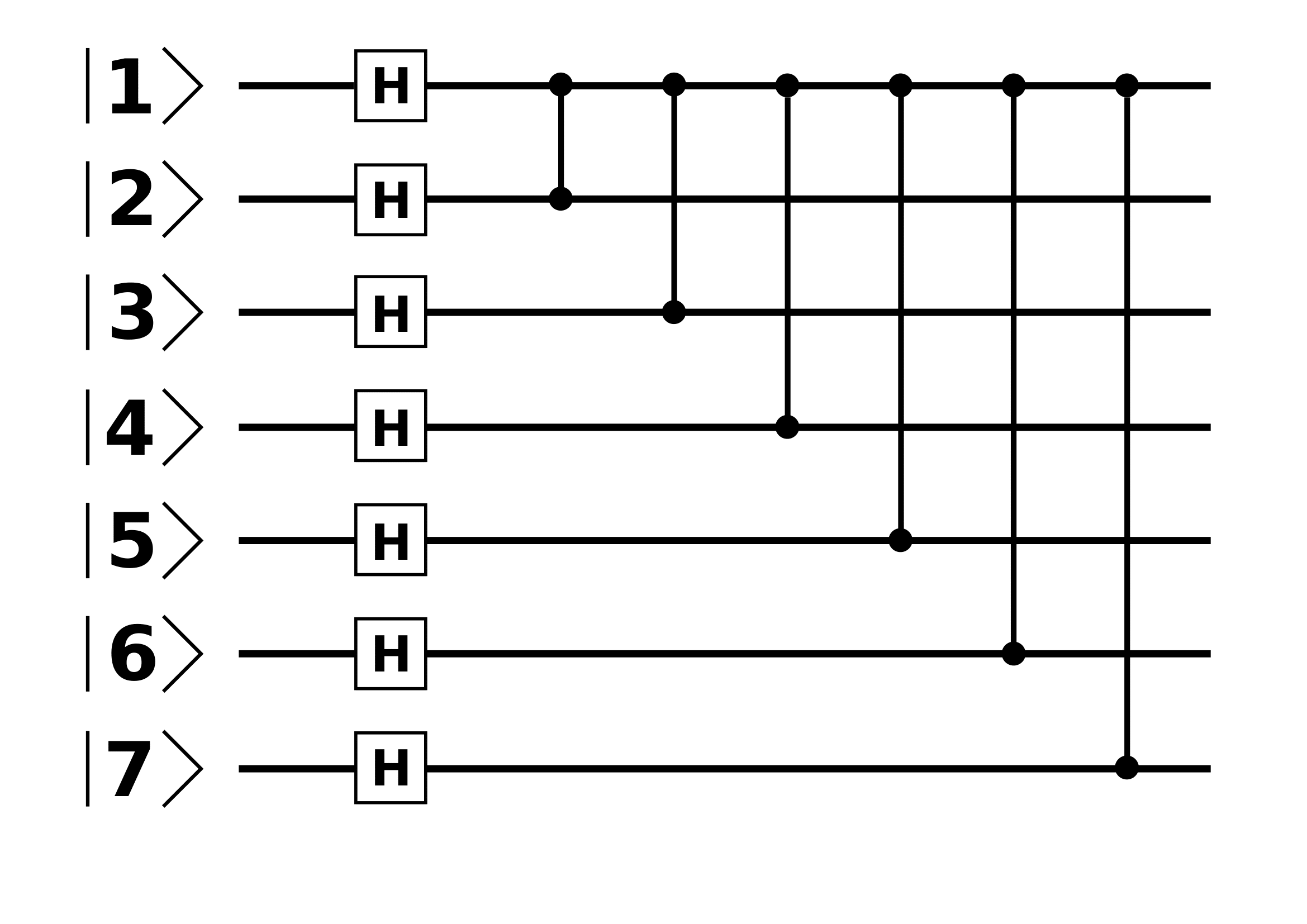} }}
\centering\caption{\label{fig:CHZgraph}(a) A star state with seven vertices, or $\left|V \right|=7$. A sole vertex (the central one, $1$) connects the whole graph. An algebraic representation is given in Eq. \eqref{eq:GHZ7state}. (b) Preparing a register of seven qubits in the star state, represented as a graph in the above picture. The qubits in the register, before to build the circuit, lie in the $\ket{0}$ state. In the first place, a Hadamard operator acts on every qubit, so to morph their state from $\ket{0}$ to $\ket{+}$, then a set of $\hat{CZ}$ gates is applied, using the first qubit as control and all the others as target.}
\end{figure}

A graph $G[V,E]$ is a triple~\cite{west2001introduction} consisting of a vertex set $V(G) = \{1, ..., N\}$, an edge set $E(G)$, and a relation that associates with each edge two vertices (not necessarily distinct) called its endpoints. The elements from $V(G)$ are called vertices of the graph $G$, the elements from $E(G)$ its edges. From now on, the edges $e_i \in E(G)$ will be referred to by specifying their endpoints, so that an edge $e_i$ with endpoints $a$, $b$ in $V(G)$ will be labelled as $(a,b)$. With such notation, it also follows that $E(G) \subset [V(G)]^2$~\cite{hein2006entanglement}. A loop is an edge whose endpoints are equal. Multiple edges are edges having the same pair of endpoints. A simple graph is a graph having no loops or multiple edges. An example of simple graph is given in Fig. \ref{fig:CHZgraph}, where the vertices are the circles labelled from $1$ to $7$ (i.e. $N=7$), and the edges the connections between them. The connection between $1$ and $2$, for instance, is the $(1, 2)$ edge.

An equivalence relation between the edges can be established:
\begin{equation}
\label{eq:equivrel}
    (a,b) \sim (c, d) \Leftrightarrow a=d \wedge b = c
\end{equation}
With such a formulation, it is possible to state that $(a,b) \sim (b,a)$. The neighborhood $N_a$ of a vertex $a$ is the collection of all the vertices connected to $a$ through an edge:
\begin{equation}
\label{eq:NNa}
    N_a = \{ b\in V | (a,b) \in E \}
\end{equation}
Edges provide a formal representation of quantum entanglement, therefore the graph states are concerned only with simple graphs.

\begin{figure}
\subfloat[\label{subfig:ring}]{{\includegraphics[width=3cm]{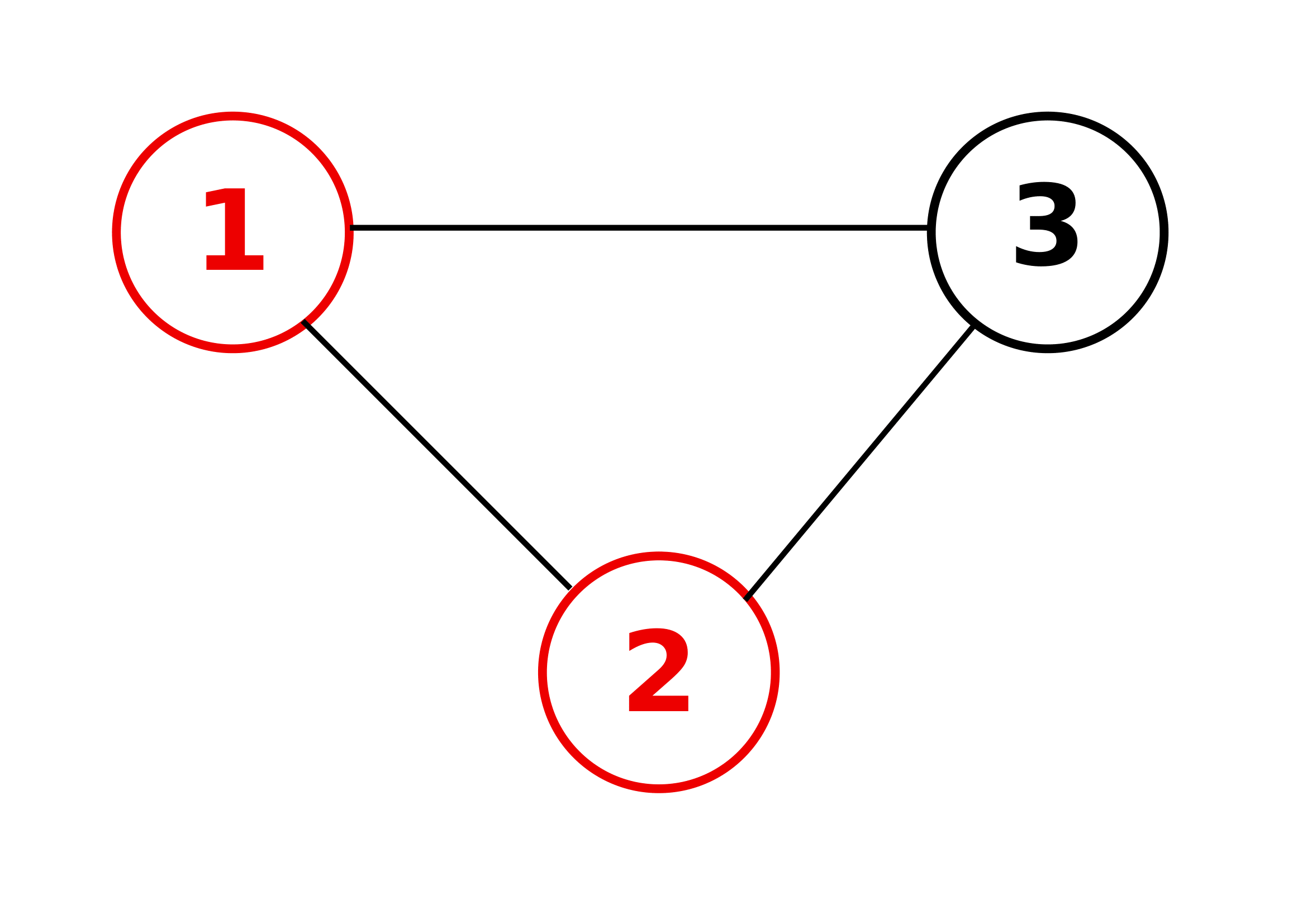} }}
\quad
\subfloat[\label{subfig:GHZhub}]{{\includegraphics[width=5.53cm]{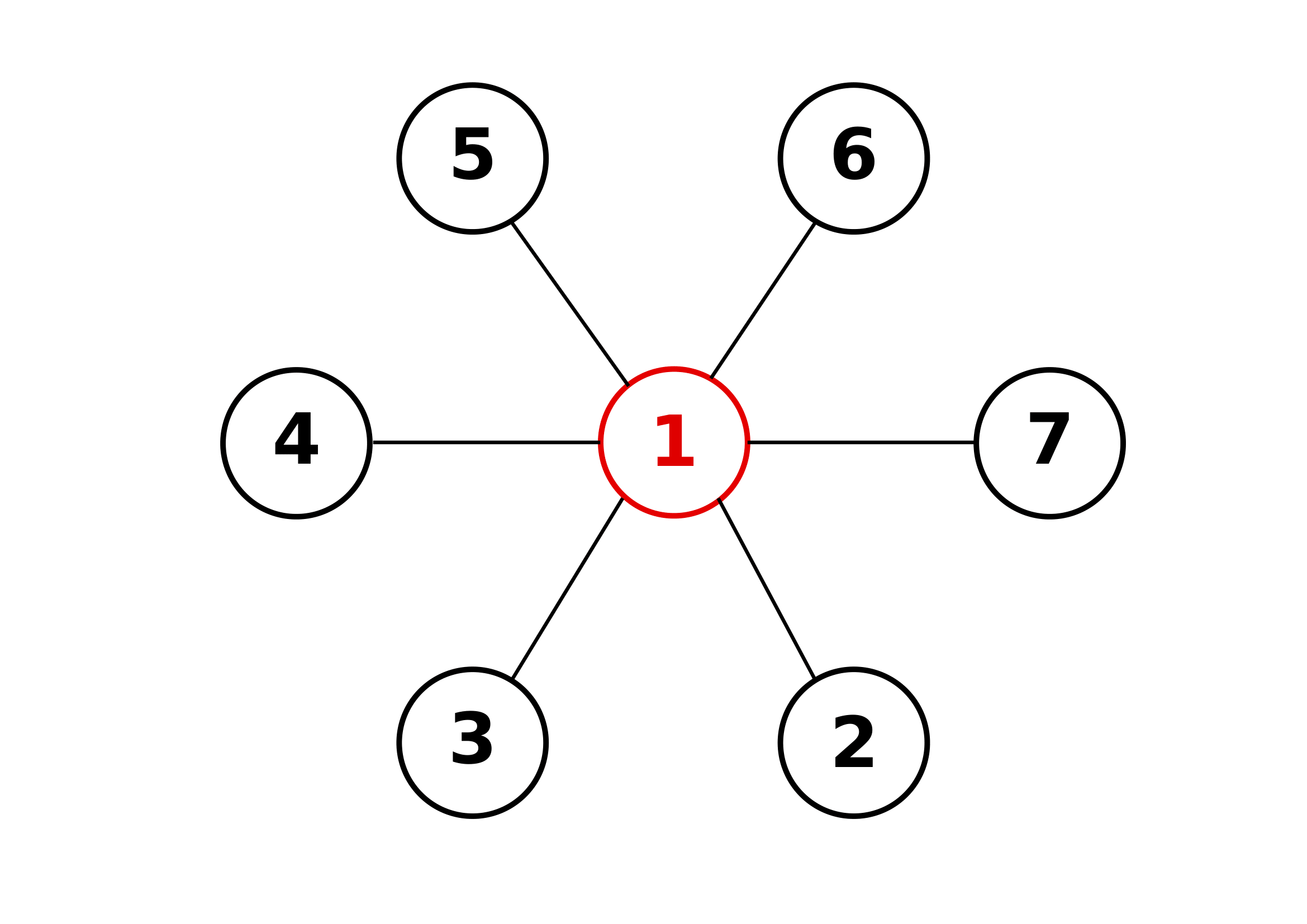} }}
\centering\caption{\label{fig:Hubs}(a) A graph picturing a ring with three vertices. The nodes in red $(1,2)$ represent the hubs of the graph. Such choice is arbitrary, as far as choosing $(1,3)$ or $(2,3)$ would have mapped as well all the edges of graph. As a matter of fact, such graph represents a perfectly homogeneous network. (b) The same star state from Fig. \ref{fig:CHZgraph}, this time the central node is highlighted in red, in order to underline it serves as hub. The graph for any star state represents a highly heterogeneous network, where a unique node keeps all the edge connections in the graph.}
\end{figure}

In network theory, a hub is defined as a highly connected vertex, i.e. a nodal vertex which is linked to several other vertices~\cite{caldarelli2012networks}. Networks which display a robust number of hubs can be depicted by heterogeneous graphs, i.e. graphs whose connections are quite asymmetric, as far as few vertices connect the whole graph. The network depicted in Fig. \ref{subfig:GHZ} is highly heterogeneous, as far as a sole hub connects all the other nodes. On the other hand, the ring graphs, such as in Fig. \ref{subfig:ring}, are examples of perfectly homogeneous networks, where every vertex shares the same number of connections. Referring to the term hub, we mean those nodes whose connections span the whole $E$ space, i.e. which collect all the edges of the network. In Fig. \ref{subfig:GHZhub} all the connections are spanned from the central vertex, which can be righteously promoted to hub. With regard to Eq. \eqref{eq:NNa}, $N_1=E$, i.e. the neighborhood of the first vertex spans all the edges in the graph.

Hubs can be applied to describe graph states. Given a graph $G[V,E]$, there exists a subset $\mathcal{B} \subset V$ such that
\begin{equation}
\label{eq:HubsSet}
    \bigcup_{a \in \mathcal{B}} N_a \supseteq E
\end{equation}
i.e. the union of the neighborhoods of all the hubs covers the entire edge space. Recalling the equivalence relation in Eq. \eqref{eq:equivrel}, it is possible to state that
\begin{equation}
\label{eq:equivEdge}
    \sfrac{\bigcup_{a \in \mathcal{B}} N_a}{\sim} = E
\end{equation}
i.e. the quotient set of the equivalence class between the edges is the edge set. In other terms, the union of all the equivalence classes of edges is the edge set itself. This feature will be crucial when building an algebraic representation in sec. (\ref{subsec:algebraic_graphStates}). To clarify, in Fig. \ref{fig:CHZgraph} the hub set $\mathcal{B}$ is given by $1$. It means that its neighborhood equals the whole $E$ set:
\begin{equation}
    N_1 = \{ (1,2), (1,3), (1,4), (1,5), (1,6), (1,7) \} = E
\end{equation}
However, taking the example in Fig. \ref{subfig:ring} and promoting the first two vertex to hubs, the hub $\mathcal{B}=\{1,2\}$ set now covers, but does not equal the edge set:
\begin{equation}
    \bigcup_{a \in \{1,2\}} N_a = \{ (1,2), (1,3), (2,1), (2,3)  \} \supset E
\end{equation}
To equal the edge set, it is mandatory to take only the representatives for the equivalence relation in Eq. \eqref{eq:equivrel}, which means $(2,1)$ or $(1,2)$ must be erased:
\begin{equation}
    \sfrac{\bigcup_{a \in \{1,2\}} N_a}{\sim} = \{ (1,2), (1,3), (2,3) \} = E
\end{equation}
%

\section{Algebraic representation}
\label{subsec:algebraic_graphStates}

The graph state is the translation of a graph $G[V,E]$ in a Hilbert space. In this graphical representation, a vertex $a$ stands for a qubit, while the edges represent the entanglement between two qubits. To produce such an entangled state, an example is given in Fig. $\ref{subfig:CHZregister}$ for the star state. The graph state and the qubit register, as shown in Fig. $\ref{fig:CHZgraph}$, are in fact two equivalent representations of the graph state~\cite{rossi2013quantum}.\\
In the first place, given a register of $n$ qubit, let's prepare them in the $\ket{+}$ state. Conventionally, the qubits in quantum circuits are initialized in the $\ket{0}$ state, thus apply a Hadamard gate to switch them to the $\ket{+}$ state. The global state $\ket{\Psi}$ now reads
\begin{equation}
\label{eq:plus_state}
    \ket{\Psi}_7 = \ket{+}^{\otimes n}
\end{equation}
Secondly, to entangle two qubits $i$ and $j$, one applies a $\hat{CZ}_{ij}$ gate on the state $\ket{\Psi}$. The $\hat{CZ}_{ij}$, i.e. the controlled z-gate (or phase gate), is defined as follows:
\begin{equation}
\label{eq:CZketdef}
    \hat{CZ}_{ij} = \ket{0}_i \bra{0}_i \otimes \hat{I}_j + \ket{1}_i\bra{1}_i \otimes \hat{Z}_j 
\end{equation}
Any $\hat{CZ}_{ij}$ commutes each other with, more specifically $\left[ \hat{CZ}_{ij} \, ; \, \hat{CZ}_{km} \right] = 0 \; \forall i,j,k,m$, as all diagonal matrices commute each other with. As a consequence, the global state $\ket{\Psi}$ can be entangled without any worry about how to apply the sequence of $\hat{CZ}_{ij}$. The action of the controlled phase operator can be summed up:
\begin{equation}
\label{eq:CZx}
\begin{cases}
    \hat{CZ} \ket{0+} = \ket{0+}\\
    \hat{CZ} \ket{0-} = \ket{0-}\\
    \hat{CZ} \ket{1+} = \ket{1-}\\
    \hat{CZ} \ket{1-} = \ket{1+}
\end{cases} \quad
\begin{cases}
    \hat{CZ} \ket{00} = \ket{00}\\
    \hat{CZ} \ket{01} = \ket{01}\\
    \hat{CZ} \ket{10} = \ket{10}\\
    \hat{CZ} \ket{11} = -\ket{11}
\end{cases}
\end{equation}
To build the star state in Fig. \ref{fig:CHZgraph}, first prepare the register in the $\ket{+}$ state, in eq. \eqref{eq:plus_state}. However, in order to apply the $\hat{CZ}$ gates, it is smart to select the fewest control qubits, as they have to be rewritten as $\ket{0}+\ket{1}$. For the star state, selecting the first qubit as the control means to rewrite the overall state as
\begin{equation}
    \ket{\Psi}_7 = \frac{\ket{0++++++}+\ket{1++++++}}{\sqrt{2}}
\end{equation}
then apply the control-z gates:
\begin{equation}
\label{eq:GHZ7state}
    \bigotimes_{j=2}^7 \hat{CZ}_{1j} \ket{\Psi}_7 = \frac{\ket{0++++++}+\ket{1------}}{\sqrt{2}} 
\end{equation}
to obtain the corresponding $\ket{\Psi_{star}}_7$ state.

\subsection{Hubs and control qubits}

As seen in the previous sections, the quotient set on the hubs in the graph gets all the edges, i.e. all the entanglements between the qubits. Thus, in order to write down the ket for a graph state, it could be useful to promote the hubs of a graph to the control qubits, when applying the $\hat{CZ}_{ij}$ operators on the starting $\ket{+}^{\otimes n}$ register, as pursued for the star state in Eq. \eqref{eq:GHZ7state}. In such case, the central hub was selected as control qubit, from which it has been possible to track down all the entanglements for the quantum state, i.e. to apply the $\hat{CZ}_{1j}$ operators. The general way to rewrite such operation is the following:
\begin{equation}
\label{eq:preparingState}
    \bigotimes_{(i,j) \in \sfrac{\bigcup_{a \in \mathcal{B}} N_a}{\sim}} \hat{CZ}_{ij} \ket{+}^{\otimes n}
\end{equation}
where $\sfrac{\bigcup_{a \in \mathcal{B}} N_a}{\sim}$ is the quotient set defined in Eq. \eqref{eq:equivEdge}, and $i$, $j$ are a couples of vertices. In practice, in Eq. \eqref{eq:GHZ7state} the quotient space is $\{ (1,2), (1,3), (1,4), (1,5), (1,6), (1,7) \}$, and the $i$, $j$ pairs are $i=1$ $j=2$, $i=1$ $j=3$ and so on. Thus, the quotient set equals the edge space $E$, but specifying how the couples of vertices are ordered, which can prove to be an advantage when representing the graph state in its algebraic fashion. Such advantage will be clarified in sec. ($\ref{subsec:RulesToWrite}$).

Let's now consider another example i.e. the ring state in Fig. \ref{subfig:ring}. There are three options to choose the hubs, i.e. $\mathcal{B}$ can be with no difference $\{1,2\}$, $\{1,3\}$ or $\{2,3\}$. In figure, the choice is $\mathcal{B}=\{1,2\}$. In this case, the starting register $\ket{\Psi}$ is written as
\begin{equation}
\label{eq:ringstateregister}
    \ket{\Psi} = \ket{+++} = \frac{1}{2} \left( \ket{00} + \ket{01} + \ket{10} + \ket{11} \right)_{12} \ket{+}_3
\end{equation}
Let's now apply the $\hat{CZ}$ gates, taking $1$ and $2$ as control qubits and $3$ as the target one. In order to entangle $1$ with $2$, one of them must be taken as target qubit for the $\hat{CZ}_{1,2}$ operator. The quotient set on the hubs, for the ring graph, is
\begin{equation}
    \sfrac{\bigcup_{a \in \mathcal{B}} N_a}{\sim} = \{ (1,2), (1,3), (2,3) \}
\end{equation}
thus the $\hat{CZ}$ operators to apply on the state in Eq. \eqref{eq:ringstateregister} will be
    $\hat{CZ}_{1,2} \hat{CZ}_{1,3} \hat{CZ}_{2,3}$
and the final state $\ket{\Psi_{ring}}_3$, applying the rules from eqs. $\eqref{eq:CZx}$, will be
\begin{equation}
    \ket{\Psi_{ring}}_3 = \frac{1}{2} \left( \ket{00+} + \ket{01-} + \ket{10-} - \ket{11+} \right)
\end{equation}
%

\section{Stabilizers formalism}

In this section, the formalism for the stabilizer operators to manipulate more efficiently the graph states is introduced, based on the usual definition of the n-qubit Pauli group $\mathcal{P}_n$ \cite{fujii2015quantum}.
%
%
%
%
A subgroup from the Pauli's one is the stabilizer group $\mathcal{S}_n$, endowed with the property to be abelian~\cite{rodriguez2021efficient}:
\begin{equation}
\label{eq:StabGroup}
    \mathcal{S}_n = \{ \hat{S}_i \in \mathcal{P}_n \; | \; \hat{S}_i^\dagger = \hat{S}_i, [\hat{S}_i, \hat{S}_j] = 0 \; \forall i,j \}
\end{equation}
Keeping on with the $\mathcal{P}_2$ example, its stabilizer subgroup is given by $\mathcal{S}_2 = \{ \hat{I}_1 \hat{I}_2, \hat{X}_1 \hat{X}_2, -\hat{Y}_1 \hat{Y}_2, \hat{Z}_3 \hat{Z}_3 \}$. The elements in $\mathcal{S}_2$ form the biggest abelian subgroup for $\mathcal{P}_2$. However, $-\hat{Y}_1 \hat{Y}_2$ can be obtained from the composition of $\hat{X}_1 \hat{X}_2$ and $\hat{Z}_1 \hat{Z}_2$ since $\sigma_i \sigma_j = \delta_{ij} I + i \epsilon_{ijk} \sigma_k$:
\begin{equation}
    \hat{X}_1 \hat{X}_2 \circ \hat{Z}_1 \hat{Z}_2 = ( \hat{X}_1 \hat{Z}_1 ) (\hat{X}_2 \hat{Z}_2 ) = - \hat{Y}_1 \hat{Y}_2
\end{equation}
Eventually, the double identity can be easily achieved by both $(\hat{X}_1 \hat{X}_2)^2$ or $(\hat{Z}_1 \hat{Z}_2)^2$. Thus the $\mathcal{S}_2$ group can be written in terms of its generators: $\mathcal{S}_2 = \langle \hat{X}_1 \hat{X}_2, \hat{Z}_1 \hat{Z}_2 \rangle$.


As far as any stabilizer group forms an abelian group of commuting Hermitian operators, such set of operators is able to span a space of eigenstates, the stabilizer states~\cite{anders2006fast,pfister2019continuous}.
%
%
For instance, the generators for a $GHZ$ state are given by
\begin{equation}
\label{eq:SnStabilizers}
    \mathcal{S}_n = \langle \hat{Z}_1 \hat{Z}_2; \hat{Z}_2 \hat{Z}_3; ...; \hat{Z}_{n-1} \hat{Z}_n; \bigotimes_{i=1}^n \hat{X}_i \rangle
\end{equation}
where a $GHZ$ state, with $n$ vertices, is given by
\begin{equation}
\label{eq:GHZ_n}
    \ket{GHZ}_n = \frac{1}{\sqrt{2}}\left(\ket{0}^{\otimes n} + \ket{1}^{\otimes n}\right)
\end{equation}
Indeed, $\ket{GHZ}_n$ is an eigenvector for all of the operators in Eq. \eqref{eq:SnStabilizers}, and the number of operators matches the dimension of $\ket{GHZ}_n$. Moreover, they commute each other with.

The expression of the stabilizer group via its generators is not unique. In fact, it is possible to transform a stabilizer state $\ket{\Psi}$ via an unitary $\hat{U}$, and thus the stabilizers themselves:
\begin{equation}
\label{eq:Clifford}
    \hat{U} \ket{\Psi}  = \hat{U} \hat{S}_i \ket{\Psi} = \hat{U} \hat{S}_i \hat{U}^\dagger \hat{U} \ket{\Psi} = \hat{S}_i' \hat{U} \ket{\Psi}
\end{equation}
Such transformations, which affect both the stabilizer states and operators, are called Clifford operations. The Clifford group is generated by three elementary gates, i.e. the Hadamard gate, the $\pi/4$ phase rotation and, by choice, the CNOT or the CZ gate~\cite{anders2006fast}.

\subsection{Stabilizer states and graph states}

Given an $n$-qubit system, its initial register $\ket{\Psi}$ is prepared in the state $\ket{\Psi} = \ket{+}^{\otimes n}$.
Such state shares a set of stabilizers, which are $\langle \hat{X}_1, \hat{X}_2, ..., \hat{X}_n \rangle$.
Any Clifford operation maps a stabilizer state into a different representation, as well for its stabilizers. The purpose is now to map such stabilizer states into graph states via a set of Clifford operations. Applying a $\hat{CZ}_{ij}$ gate on the $\ket{\Psi}$ state, it is possible to set the transformation for the $\hat{X}_i$, $\hat{X}_j$ recalling the Clifford operation in Eq. \eqref{eq:Clifford}:
\begin{equation}
    \hat{CZ}_{ij} \ket{\Psi} = \hat{CZ}_{ij} \hat{X}_i \hat{CZ}_{ij} \hat{CZ}_{ij} \ket{\Psi}
\end{equation}
as far as $\hat{CZ}_{ij}^\dagger = \hat{CZ}_{ij}$, as it can be check from its definitions from Eq. \eqref{eq:CZketdef}. It is easy to prove the following rules~\cite{danos2007measurement, pius2010automatic}:
\begin{equation}
\label{eq:CzX}
    \begin{cases}
    \hat{CZ}_{ij} \, \hat{X}_i = \hat{X}_i \hat{Z}_j \, \hat{CZ}_{ij} \\
    \hat{CZ}_{ij} \, \hat{X}_j = \hat{Z}_i \hat{X}_j \, \hat{CZ}_{ij} 
    \end{cases}
\end{equation}
Generalizing to $\hat{U}=\bigotimes_{(i,j) \in E} \hat{CZ}_{ij}$ Clifford operation, it follows straightforward that a single $\hat{X}_i$ transforms into a stabilizer $\hat{K}_i$:
\begin{equation}
\label{eq:clusterOp}
    \hat{U} \hat{X}_i \hat{U}^\dagger = \hat{X}_i \bigotimes_{j \in \langle i \rangle} \hat{Z}_j = \hat{K}_i
\end{equation}
where $\langle i \rangle$ stands for the neighborhood of the $i$ vertex. We will refer to the $\hat{K}_i$ operators also as cluster operators, or cluster stabilizers.

\subsection{Stabilizers of a graph state}
\label{subsec:stabilizer_rule}

From any graph representation, it is possible to express the stabilizers of the system. When the register of qubits is prepared in the $\ket{+}^{\otimes n}$ state, all the vertices are disconnected, which means that the stabilizers are given by
\begin{equation}
\label{eq:Xstabilizers}
    \langle \hat{X}_1; \hat{X}_2; ...; \hat{X}_i; ...; \hat{X}_j; ...; \hat{X}_n \rangle
\end{equation}
as far as the $\ket{+}^{\otimes n}$ state is stabilized by the operators above, i.e. it is an eigenvector for every $\hat{X}_i$ observable with eigenvalue $+1$. When applying a $\hat{CZ}_{ij}$ gate on the $\ket{+}^{\otimes n}$ register, such transformation acts as a Clifford operation for the $\hat{X}_i$ and $\hat{X}_j$ operators themselves. Calling back the Eq. \eqref{eq:clusterOp}, the stabilizers in Eq. \eqref{eq:Xstabilizers}, after a $\hat{CZ}_{ij}$ Clifford operation, become
\begin{equation}
    \langle \hat{X}_1; \hat{X}_2; ...; \hat{X}_i \hat{Z}_j; ...; \hat{Z}_i \hat{X}_j; ... \hat{X}_n \rangle
\end{equation}
i.e. for each pair $i$, $j$ of entangled vertices, a $\hat{Z}_i$ operator is attached to the respective $\hat{X}_j$ operator and vice versa. As far as the entanglement is represented by an edge in the graph representation, to build the stabilizers for the system is mandatory to account (i) the label $i$ for the qubit, which implies a $\hat{X}_i$ operator;    (ii) the number of edges linked to the qubit itself, such that a $\hat{Z}_j$ operator will be attached for each link to the $j$-th qubit; (iii) repeat the same procedure for each qubit in the register/each vertex on the graph.

In the example of Fig. \ref{fig:CHZgraph}, the first qubit is linked to all the qubits in the register, giving the stabilizer:
\begin{equation}
    \hat{X}_1 \rightarrow \hat{X}_1 \hat{Z}_2 \hat{Z}_3 \hat{Z}_4 \hat{Z}_5 \hat{Z}_6 \hat{Z}_7 = \hat{X}_1 \bigotimes_{i=2}^7 \hat{Z}_i
\end{equation}
while the stabilizers for all the other $i$-th qubits transform as $\hat{X}_i \rightarrow \hat{Z}_1 \hat{X}_i$. Eventually the overall stabilizers are given by
\begin{equation}
\label{eq:3stabilizers}
    \langle \hat{X}_1 \otimes_{i=2}^7 \hat{Z}_i; \hat{Z}_1 \hat{X}_2; \hat{Z}_1 \hat{X}_3; \hat{Z}_1 \hat{X}_4; \hat{Z}_1 \hat{X}_5; \hat{Z}_1 \hat{X}_6; \hat{Z}_1 \hat{X}_7  \rangle
\end{equation}

\section{Expression of the graph state via the hub stabilizers}
\label{subsec:RulesToWrite}

We now turn to the algorithm for efficiently expressing the ket of the graph state via the $\hat{K}_i$ stabilizers from the $i$-th hubs in $\mathcal{B}$. The first step consists of the preparation of a register for $n$ qubits, referring to the scheme in Fig. \ref{fig:CHZgraph} and Eq. \eqref{eq:preparingState}:
\begin{equation}
\label{eq:allCZ}
    \bigotimes_{(i,j) \in \sfrac{\bigcup_{a \in \mathcal{B}} N_a}{\sim}} \hat{CZ}_{ij} \ket{+}^{\otimes n}
\end{equation}
%
\textbf{Theorem} The application of the set of control gates $\hat{CZ}_{ij}$ in Eq. \eqref{eq:allCZ}, over the initial state $\ket{+}^{\otimes n}$, can be rewritten in terms of the cluster operators of the hubs:
\begin{equation}
\label{eq:CZisK}
    \bigotimes_{(i,j) \in \sfrac{\bigcup_{a \in \mathcal{B}} N_a}{\sim}} \hat{CZ}_{ij} \ket{+}^{\otimes n} = \bigotimes_{i \in  \mathcal{B} } \left( \frac{\hat{\mathbb{I}} + \hat{K}_i}{\sqrt{2}} \right) \ket{0}_i \ket{+}^{\otimes (n-\left| \mathcal{B} \right|)}
\end{equation}
$\left| \mathcal{B}\right|$ being the cardinality of $\mathcal{B}$.\\
\textbf{Proof} The $\ket{+}^{\otimes n}$ state can be decomposed as follows:
\begin{equation}
    \left[ \frac{1}{\sqrt{2}} \left( \ket{0} + \ket{1} \right) \right]^{\otimes \left| \mathcal{B} \right|} \ket{+}^{\otimes (n-\left| \mathcal{B} \right|)}
\end{equation}
The $\ket{0} + \ket{1}$ state can be rewritten as $[\hat{\mathbb{I}} + \hat{X}] \ket{0}$:
\begin{equation}
    \bigotimes_{i \in \mathcal{B} } \left[\frac{1}{\sqrt{2}} \left( \hat{\mathbb{I}} + \hat{X}_i \right) \ket{0}_i \right] \ket{+}^{\otimes (n-\left| \mathcal{B} \right|)}
\end{equation}
As the $\hat{CZ}$ operators in Eq. \eqref{eq:allCZ} cover all the edges, applying them  on the left member as in Eq. \eqref{eq:CZisK}, and following the rules in Eqs. \eqref{eq:CzX} the graph state becomes
\begin{equation}
    \bigotimes_{i \in \mathcal{B}} \left[ \frac{1}{\sqrt{2}} \left( \hat{\mathbb{I}} + \hat{X}_i \bigotimes_{j \in \langle i\rangle} \hat{Z}_j \right) \ket{0}_i \right] \ket{+}^{\otimes (n-\left|\mathcal{B}\right|)}
\end{equation}
where $\langle i \rangle$ is the set of nearest neighbors for the $i$-th vertex. From Eq. \eqref{eq:CZx} the $\hat{CZ}_{ij}$ operator does not affect any $\ket{0}_i \ket{+}_j$ state.
The $\hat{X}_i \otimes_{j \in \langle i \rangle} \hat{Z}_j$ can be recognized as the cluster operator $\hat{K}_i$ in the form of Eq. \eqref{eq:clusterOp} for the hubs. Thus the graph state can be rewritten as
\begin{equation}
    \bigotimes_{i \in \mathcal{B}} \left( \frac{\hat{\mathbb{I}} + \hat{K}_i}{\sqrt{2}} \right) \ket{0}_i \ket{+}^{\otimes (n-\left| \mathcal{B} \right|)}
\end{equation}
which ends the proof. $\blacksquare$\\
The same state can be written as
\begin{equation}
    \frac{1}{\sqrt{2^{b}}} \left( \hat{\mathbb{I}} + \hat{K}_{1} \right) ... \left( \hat{\mathbb{I}} + \hat{K}_{b} \right) \ket{0}^{b} \ket{+}^{\otimes (n-b)}
\end{equation}
where we rewrote $\left|\mathcal{B}\right|$ as $b$. As the normalization constant scales as $1/\sqrt{2^{ b }}$, we have a hint to foresee that the number of operators scales as $2^{\left| \mathcal{B} \right|}$, where $\left| \mathcal{B} \right|$ is the number of hubs (and thus of control qubits). For this reason, choosing correctly the hubs of a network (i.e. the control qubits of the system) can ease the calculations by far. We remark that the expansion of the product of the operator terms is expressed by
\begin{equation}
\begin{split}
    S_{\mathcal{B}}[\hat{K}] = \left( \hat{\mathbb{I}} + \hat{K}_{1} \right) ... \left( \hat{\mathbb{I}} + \hat{K}_{b} \right) = \\
    \hat{\mathbb{I}} + \hat{K}_1 + \hat{K}_2 + ... + \hat{K}_b + \hat{K}_1 \hat{K}_2 + ... + \hat{K}_{b-1} \hat{K}_b + \\
     + \hat{K}_1 \hat{K}_2 \hat{K}_3 + ... + \hat{K}_{b-2} \hat{K}_{b-1} \hat{K}_b  + \hat{K}_1 \hat{K}_2 \hat{K}_3 \hat{K}_4 +  \\
    + ... + \hat{K}_{b-3} \hat{K}_{b-2} \hat{K}_{b-1} \hat{K}_b + ... + \hat{K}_1 \hat{K}_2 \hat{K}_3 ... \hat{K}_{b-1} \hat{K}_b
\end{split}
\end{equation}
which resembles the $\mathcal{T}$-ordered exponential, with respect to the label $i$ instead of $t_i$, truncated at the $b$-th order. The final expression of any graph state with $b$ hubs is
\begin{equation}
\label{eq:Texp}
     S_\mathcal{B}[\hat{K}] \frac{\ket{0}^{b} \ket{+}^{\otimes (n-b)}}{\sqrt{2^{b}}}
\end{equation}
Such formulation provides an algorithm to express any graph state just selecting its hubs. This work is now completed with two examples, based on the formula given by Eq. $\eqref{eq:Texp}$.



\subsection{First example: star topology}

Let's start from the usual star graph from Fig. \ref{subfig:GHZ}, which we have already dealt with in sec. (\ref{subsec:algebraic_graphStates}) to describe its algebraic representation, and in sec. (\ref{subsec:stabilizer_rule}) to write down its stabilizers, described in Eq. \eqref{eq:3stabilizers}. The $\mathcal{T}$ ordered product for the $\hat{K}_1$ operator will thus be
\begin{equation}
    S_{\{1\}}[\hat{K}] = \hat{\mathbb{I}} + \hat{X}_1 \hat{Z}_2 \hat{Z}_3 \hat{Z}_4 \hat{Z}_5 \hat{Z}_6 \hat{Z}_7
\end{equation}
and the corresponding state ``at time $t_0$'' is
\begin{equation}
    \frac{\ket{0}\ket{+}^{\otimes 6}}{\sqrt{2}}
\end{equation}
Applying the $S$ operator, it follows
\begin{equation}
\label{eq:star}
    S_{\{1\}}[\hat{K}] \frac{\ket{0}\ket{+}^{\otimes 6}}{\sqrt{2}} = \frac{\ket{0}\ket{+}^{\otimes 6} + \ket{1}\ket{-}^{\otimes 6}}{\sqrt{2}}
\end{equation}
Such result can be generalized for any $n \neq 7$: given a star topology, to describe the corresponding graph it suffices the stabilizer of the central vertex, reducing from $n$ to $1$ the number of stabilizers to describe the system. Furthermore, comparing Eq. \eqref{eq:star} with \eqref{eq:GHZ_n}, it is possible to see that the $GHZ$ and the star states are equivalent up to a set of $\hat{H}^{\otimes (n-1)}$ gates.

\subsection{Second example: the three-vertex ring}

As second example, take the graph state from Fig. \ref{subfig:ring} and repeat the procedure. In the first place, write down the stabilizers:
\begin{equation}
\label{eq:ringStab}
    \langle \hat{X}_1 \hat{Z}_2 \hat{Z}_3; \hat{Z}_1 \hat{X}_2 \hat{Z}_3; \hat{Z}_1 \hat{Z}_2 \hat{X}_3 \rangle
\end{equation}
In second place, select a number of hubs so that the entire $E$ space is covered. The choice from Fig. \ref{subfig:ring} proves to be a good one. The $t_0$ ket, following the rule in Eq. \eqref{eq:Texp}, would be $\ket{00+}/2$, as far as the $1$ and $2$ qubits have been promoted to hubs, which implies $\ket{+}_1$, $\ket{+}_2$ to be flipped to $\ket{0}_1$, $\ket{0}_2$. The operator $S_{\{1,2\}}$ is
\begin{equation}
    S_{\{1,2\}}[\hat{K}] =
    \hat{\mathbb{I}} + \hat{X}_1 \hat{Z}_2 \hat{Z}_3 + \hat{Z}_1 \hat{X}_2 \hat{Z}_3 + \hat{X}_1 \hat{Z}_1 \hat{Z}_2 \hat{X}_2 
\end{equation}
At last, apply $S$ onto the $t_0$ state:
\begin{equation}
    S_\mathcal{B}[\hat{K}] \frac{\ket{00+}}{2} = \frac{\ket{00+} + \ket{10-} + \ket{01-} - \ket{11+} }{2}
\end{equation}
As before, such result can be generalized for any $n > 3$: given a ring topology, to describe the corresponding graph it suffices to take half of the vertices (for an even number of nodes) or half plus one (for an odd number of nodes), reducing from $n$ to $\ceil{\frac{n}{2}}$ the number of stabilizers to describe the system.

\section{Conclusions}
To conclude, by starting from special hub nodes, we demonstrated how to efficiently express a graph state through the generators of the stabilizer group. We have provided the expression on a ring and on a star topology, respectively. We have demonstrated that the graph states can be generated by a subgroup of the stabilizer group, so to manipulate the graph states with a reduced number of stabilizers. For the ring topology, which is completely homogeneous, the number of stabilizers required diminishes from $n$ to $\ceil{\frac{n}{2}}$, while for the star topology, which displays highly heterogeneous graphs, the number of stabilizers can be reduced from $n$ to $1$. Moreover, the same algorithm can be extended to any class of states equivalent to a graph state via a set of Clifford operations, as for the $GHZ$ states.\\
\textit{Acknowledgements} This research has been partially funded by Leonardo SPA

\bibliographystyle{IEEEtran} 
\bibliography{references} 

\begin{thebibliography}{10}
\providecommand{\url}[1]{#1}
\csname url@samestyle\endcsname
\providecommand{\newblock}{\relax}
\providecommand{\bibinfo}[2]{#2}
\providecommand{\BIBentrySTDinterwordspacing}{\spaceskip=0pt\relax}
\providecommand{\BIBentryALTinterwordstretchfactor}{4}
\providecommand{\BIBentryALTinterwordspacing}{\spaceskip=\fontdimen2\font plus
\BIBentryALTinterwordstretchfactor\fontdimen3\font minus
  \fontdimen4\font\relax}
\providecommand{\BIBforeignlanguage}[2]{{%
\expandafter\ifx\csname l@#1\endcsname\relax
\typeout{** WARNING: IEEEtran.bst: No hyphenation pattern has been}%
\typeout{** loaded for the language `#1'. Using the pattern for}%
\typeout{** the default language instead.}%
\else
\language=\csname l@#1\endcsname
\fi
#2}}
\providecommand{\BIBdecl}{\relax}
\BIBdecl

\bibitem{briegel2009measurement}
H.~J. Briegel, D.~E. Browne, W.~D{\"u}r, R.~Raussendorf, and M.~Van~den Nest,
  ``Measurement-based quantum computation,'' \emph{Nature Physics}, vol.~5,
  no.~1, pp. 19--26, 2009.

\bibitem{adcock2019programmable}
J.~C. Adcock, C.~Vigliar, R.~Santagati, J.~W. Silverstone, and M.~G. Thompson,
  ``\href{https://www.nature.com/articles/s41467-019-11489-y}{Programmable
  four-photon graph states on a silicon chip},'' \emph{Nature communications},
  vol.~10, no.~1, pp. 1--6, 2019.

\bibitem{scott2022timing}
J.~R. Scott and K.~C. Balram,
  ``\href{https://ieeexplore.ieee.org/stamp/stamp.jsp?arnumber=9779088}{Timing
  constraints imposed by classical digital control systems on photonic
  implementations of measurement-based quantum computing},'' \emph{IEEE
  Transactions on Quantum Engineering}, vol.~3, pp. 1--20, 2022.

\bibitem{wang2020integrated}
J.~Wang, F.~Sciarrino, A.~Laing, and M.~G. Thompson,
  ``\href{https://www.nature.com/articles/s41566-019-0532-1}{Integrated
  photonic quantum technologies},'' \emph{Nature Photonics}, vol.~14, no.~5,
  pp. 273--284, 2020.

\bibitem{albarran2018one}
F.~Albarr{\'a}n-Arriagada, G.~A. Barrios, M.~Sanz, G.~Romero, L.~Lamata,
  J.~Retamal, and E.~Solano,
  ``\href{https://journals.aps.org/pra/abstract/10.1103/PhysRevA.97.032320}{One-way
  quantum computing in superconducting circuits},'' \emph{Physical Review A},
  vol.~97, no.~3, p. 032320, 2018.

\bibitem{lanyon2013measurement}
B.~Lanyon, P.~Jurcevic, M.~Zwerger, C.~Hempel, E.~Martinez, W.~D{\"u}r,
  H.~Briegel, R.~Blatt, and C.~F. Roos,
  ``\href{https://journals.aps.org/prl/pdf/10.1103/PhysRevLett.111.210501?casa_token=xxlpjAlQZT8AAAAA\%3Av8EUxmwJaYW6McMCtumhnsqzOJapqGh9jn0i6fWq2NPxoBtGcD_Oa7hUX2X9hGCSLoKmTfU1cIOAgQ}{Measurement-based
  quantum computation with trapped ions},'' \emph{Physical review letters},
  vol. 111, no.~21, p. 210501, 2013.

\bibitem{blatt2012quantum}
R.~Blatt and C.~F. Roos,
  ``\href{https://www.nature.com/articles/nphys2252}{Quantum simulations with
  trapped ions},'' \emph{Nature Physics}, vol.~8, no.~4, pp. 277--284, 2012.

\bibitem{de2022temperature}
O.~de~S{\'a}~Neto, H.~Costa, G.~Prataviera, and M.~de~Oliveira,
  ``\href{https://www.nature.com/articles/s41598-022-10572-7}{Temperature
  estimation of a pair of trapped ions},'' \emph{Scientific Reports}, vol.~12,
  no.~1, pp. 1--15, 2022.

\bibitem{wigley2016fast}
P.~B. Wigley, P.~J. Everitt, A.~van~den Hengel, J.~W. Bastian, M.~A.
  Sooriyabandara, G.~D. McDonald, K.~S. Hardman, C.~D. Quinlivan, P.~Manju,
  C.~C. Kuhn \emph{et~al.},
  ``\href{https://www.nature.com/articles/srep25890?ref=https://githubhelp.com}{Fast
  machine-learning online optimization of ultra-cold-atom experiments},''
  \emph{Scientific reports}, vol.~6, no.~1, pp. 1--6, 2016.

\bibitem{krantz2019quantum}
P.~Krantz, M.~Kjaergaard, F.~Yan, T.~P. Orlando, S.~Gustavsson, and W.~D.
  Oliver,
  ``\href{https://aip.scitation.org/doi/full/10.1063/1.5089550?casa\_token=tPTVSahch-YAAAAA:30ZCO52ryVl1pagLNW53iapOxQb41Js-XrlGr9rKhGKXDCPfEXxmedzs0-TOJ6uENRuUAQP2aXO3jA}{A
  quantum engineer's guide to superconducting qubits},'' \emph{Applied Physics
  Reviews}, vol.~6, no.~2, p. 021318, 2019.

\bibitem{rotta2017quantum}
D.~Rotta, F.~Sebastiano, E.~Charbon, and E.~Prati,
  ``\href{https://www.nature.com/articles/s41534-017-0023-5}{Quantum
  information density scaling and qubit operation time constraints of CMOS
  silicon-based quantum computer architectures},'' \emph{npj Quantum
  Information}, vol.~3, no.~1, pp. 1--14, 2017.

\bibitem{aharonovich2016solid}
I.~Aharonovich, D.~Englund, and M.~Toth,
  ``\href{https://www.nature.com/articles/nphoton.2016.186}{Solid-state
  single-photon emitters},'' \emph{Nature Photonics}, vol.~10, no.~10, pp.
  631--641, 2016.

\bibitem{grosso2017tunable}
G.~Grosso, H.~Moon, B.~Lienhard, S.~Ali, D.~K. Efetov, M.~M. Furchi,
  P.~Jarillo-Herrero, M.~J. Ford, I.~Aharonovich, and D.~Englund,
  ``\href{https://www.nature.com/articles/s41467-017-00810-2}{Tunable and
  high-purity room temperature single-photon emission from atomic defects in
  hexagonal boron nitride},'' \emph{Nature communications}, vol.~8, no.~1, pp.
  1--8, 2017.

\bibitem{dietrich2020solid}
A.~Dietrich, M.~Doherty, I.~Aharonovich, and A.~Kubanek,
  ``\href{https://journals.aps.org/prb/pdf/10.1103/PhysRevB.101.081401}{Solid-state
  single photon source with Fourier transform limited lines at room
  temperature},'' \emph{Physical Review B}, vol. 101, no.~8, p. 081401, 2020.

\bibitem{zwerger2014hybrid}
M.~Zwerger, H.~Briegel, and W.~D{\"u}r,
  ``\href{https://www.nature.com/articles/srep05364}{"Hybrid architecture for
  encoded measurement-based quantum computation},'' \emph{Scientific reports},
  vol.~4, no.~1, pp. 1--5, 2014.

\bibitem{menicucci2014fault}
N.~C. Menicucci,
  ``\href{https://journals.aps.org/prl/pdf/10.1103/PhysRevLett.112.120504}{Fault-tolerant
  measurement-based quantum computing with continuous-variable cluster
  states},'' \emph{Physical review letters}, vol. 112, no.~12, p. 120504, 2014.

\bibitem{raussendorf2003measurement}
R.~Raussendorf, D.~E. Browne, and H.~J. Briegel, ``{Measurement-based quantum
  computation on cluster states},'' \emph{Physical review A}, vol.~68, no.~2,
  p. 022312, 2003.

\bibitem{fujii2015quantum}
K.~Fujii, \emph{\href{https://arxiv.org/pdf/1504.01444.pdf}{Quantum Computation
  with Topological Codes: from qubit to topological fault-tolerance}}.\hskip
  1em plus 0.5em minus 0.4em\relax Springer, 2015, vol.~8.

\bibitem{hein2004multiparty}
M.~Hein, J.~Eisert, and H.~J. Briegel, ``{Multiparty entanglement in graph
  states},'' \emph{Physical Review A}, vol.~69, no.~6, p. 062311, 2004.

\bibitem{adcock2020mapping}
J.~C. Adcock, S.~Morley-Short, A.~Dahlberg, and J.~W. Silverstone,
  ``\href{https://arxiv.org/pdf/1910.03969.pdf}{Mapping graph state orbits
  under local complementation},'' \emph{Quantum}, vol.~4, p. 305, 2020.

\bibitem{nikahd2015one}
E.~Nikahd, M.~Houshmand, M.~S. Zamani, and M.~Sedighi,
  ``\href{https://arxiv.org/pdf/1604.05659.pdf}{One-way quantum computer
  simulation},'' \emph{Microprocessors and Microsystems}, vol.~39, no.~3, pp.
  210--222, 2015.

\bibitem{morimae2014acausal}
T.~Morimae,
  ``\href{https://journals.aps.org/pra/pdf/10.1103/PhysRevA.90.010101}{Acausal
  measurement-based quantum computing},'' \emph{Physical Review A}, vol.~90,
  no.~1, p. 010101, 2014.

\bibitem{guhne2005bell}
O.~G{\"u}hne, G.~T{\'o}th, P.~Hyllus, and H.~J. Briegel,
  ``\href{https://journals.aps.org/prl/pdf/10.1103/PhysRevLett.95.120405?casa\_token=89Uy3e1Uv6QAAAAA\%3Aqlf9VoC2Ehx5\_vV94NSqtgh\_uwoxJ4N5EmlyfIDDGnzRTkPpeZEc2EXsZHWcEk2mOhjj6HIw8Zh3oA}{Bell
  inequalities for graph states},'' \emph{Physical review letters}, vol.~95,
  no.~12, p. 120405, 2005.

\bibitem{hein2006entanglement}
M.~Hein, W.~D{\"u}r, J.~Eisert, R.~Raussendorf, M.~Nest, and H.-J. Briegel,
  ``\href{https://arxiv.org/pdf/quant-ph/0602096.pdf}{Entanglement in graph
  states and its applications},'' \emph{arXiv preprint quant-ph/0602096}, 2006.

\bibitem{egan2020fault}
L.~Egan, D.~M. Debroy, C.~Noel, A.~Risinger, D.~Zhu, D.~Biswas, M.~Newman,
  M.~Li, K.~R. Brown, M.~Cetina \emph{et~al.},
  ``\href{https://arxiv.org/pdf/2009.11482.pdf}{Fault-tolerant operation of a
  quantum error-correction code},'' \emph{arXiv preprint arXiv:2009.11482},
  2020.

\bibitem{schlingemann2001quantum}
D.~Schlingemann and R.~F. Werner,
  ``\href{https://journals.aps.org/pra/pdf/10.1103/PhysRevA.65.012308?casa\_token=GTWSwJUZHKwAAAAA\%3AFaeAKPHnw3NNfRTx6hX4Ueq7Ef3nMKlEc-CXNVuBxEugx7X\_DDKaguMB8YXnZipgS4L3Tlr6FZ\_s}{Quantum
  error-correcting codes associated with graphs},'' \emph{Physical Review A},
  vol.~65, no.~1, p. 012308, 2001.

\bibitem{chiaverini2004realization}
J.~Chiaverini, D.~Leibfried, T.~Schaetz, M.~D. Barrett, R.~Blakestad,
  J.~Britton, W.~M. Itano, J.~D. Jost, E.~Knill, C.~Langer \emph{et~al.},
  ``\href{https://www.nature.com/articles/nature03074}{Realization of quantum
  error correction},'' \emph{Nature}, vol. 432, no. 7017, pp. 602--605, 2004.

\bibitem{devitt2013quantum}
S.~J. Devitt, W.~J. Munro, and K.~Nemoto,
  ``\href{https://iopscience.iop.org/article/10.1088/0034-4885/76/7/076001/pdf?casa\_token=jG0rypdmsB4AAAAA:RYZ0L2iVOsaof3UDTcdefACcLpWLzHE2Cxp4Hnp8BounRNgGX5PCCh-EdandEsT2xrXQ6wfWBw}{Quantum
  error correction for beginners},'' \emph{Reports on Progress in Physics},
  vol.~76, no.~7, p. 076001, 2013.

\bibitem{gottesman2002introduction}
D.~Gottesman, ``\href{https://arxiv.org/pdf/quant-ph/0004072.pdf}{An
  introduction to quantum error correction},'' in \emph{Proceedings of Symposia
  in Applied Mathematics}, vol.~58, 2002, pp. 221--236.

\bibitem{roffe2019quantum}
J.~Roffe, ``\href{https://arxiv.org/pdf/1907.11157.pdf}{Quantum error
  correction: an introductory guide},'' \emph{Contemporary Physics}, vol.~60,
  no.~3, pp. 226--245, 2019.

\bibitem{hsieh2007general}
M.-H. Hsieh, I.~Devetak, and T.~Brun,
  ``\href{https://journals.aps.org/pra/pdf/10.1103/PhysRevA.76.062313?casa\_token=FFove8Fn44gAAAAA\%3AB\_jOaWhwAYHwn14sJeBD2yKZixO6RSX1oJhsRlkK5040mdwD8avuXL1ODRmBLVCo5P4xVr\_yzcO5}{General
  entanglement-assisted quantum error-correcting codes},'' \emph{Physical
  Review A}, vol.~76, no.~6, p. 062313, 2007.

\bibitem{guenda2018constructions}
K.~Guenda, S.~Jitman, and T.~A. Gulliver,
  ``\href{https://link.springer.com/article/10.1007/s10623-017-0330-z}{Constructions
  of good entanglement-assisted quantum error correcting codes},''
  \emph{Designs, Codes and Cryptography}, vol.~86, no.~1, pp. 121--136, 2018.

\bibitem{nautrup2019optimizing}
H.~P. Nautrup, N.~Delfosse, V.~Dunjko, H.~J. Briegel, and N.~Friis,
  ``\href{https://quantum-journal.org/papers/q-2019-12-16-215/pdf/}{Optimizing
  quantum error correction codes with reinforcement learning},''
  \emph{Quantum}, vol.~3, p. 215, 2019.

\bibitem{gottesman1997stabilizer}
D.~Gottesman,
  \emph{\href{https://arxiv.org/pdf/quant-ph/9705052.pdf}{Stabilizer codes and
  quantum error correction}}.\hskip 1em plus 0.5em minus 0.4em\relax California
  Institute of Technology, 1997.

\bibitem{maronese2022quantum}
M.~Maronese, L.~Moro, L.~Rocutto, and E.~Prati,
  ``\href{https://arxiv.org/pdf/2112.00187.pdf}{Quantum compiling},'' in
  \emph{Quantum Computing Environments}.\hskip 1em plus 0.5em minus 0.4em\relax
  Springer, 2022, pp. 39--74.

\bibitem{maronese2022quantum2}
M.~Maronese, C.~Destri, and E.~Prati,
  ``\href{https://link.springer.com/article/10.1007/s11128-022-03466-0}{Quantum
  activation functions for quantum neural networks},'' \emph{Quantum
  Information Processing}, vol.~21, no.~4, pp. 1--24, 2022.

\bibitem{west2001introduction}
D.~B. West \emph{et~al.},
  \emph{\href{https://ia801204.us.archive.org/35/items/igt_west/igt_west_text.pdf}{Introduction
  to graph theory}}.\hskip 1em plus 0.5em minus 0.4em\relax Prentice hall Upper
  Saddle River, 2001, vol.~2.

\bibitem{caldarelli2012networks}
G.~Caldarelli and M.~Catanzaro, \emph{Networks: A very short
  introduction}.\hskip 1em plus 0.5em minus 0.4em\relax Oxford University
  Press, 2012, vol. 335.

\bibitem{rossi2013quantum}
M.~Rossi, M.~Huber, D.~Bru{\ss}, and C.~Macchiavello,
  ``\href{https://iopscience.iop.org/article/10.1088/1367-2630/15/11/113022/pdf}{Quantum
  hypergraph states},'' \emph{New Journal of Physics}, vol.~15, no.~11, p.
  113022, 2013.

\bibitem{rodriguez2021efficient}
A.~Rodriguez-Blanco, A.~Bermudez, M.~M{\"u}ller, and F.~Shahandeh,
  ``\href{https://journals.aps.org/prxquantum/pdf/10.1103/PRXQuantum.2.020304}{Efficient
  and robust certification of genuine multipartite entanglement in noisy
  quantum error correction circuits},'' \emph{PRX quantum}, vol.~2, no.~2, p.
  020304, 2021.

\bibitem{anders2006fast}
S.~Anders and H.~J. Briegel,
  ``\href{https://arxiv.org/pdf/quant-ph/0504117.pdf}{Fast simulation of
  stabilizer circuits using a graph-state representation},'' \emph{Physical
  Review A}, vol.~73, no.~2, p. 022334, 2006.

\bibitem{pfister2019continuous}
O.~Pfister,
  ``\href{https://iopscience.iop.org/article/10.1088/1361-6455/ab526f/pdf}{Continuous-variable
  quantum computing in the quantum optical frequency comb},'' \emph{Journal of
  Physics B: Atomic, Molecular and Optical Physics}, vol.~53, no.~1, p. 012001,
  2019.

\bibitem{danos2007measurement}
V.~Danos, E.~Kashefi, and P.~Panangaden,
  ``\href{https://dl.acm.org/doi/pdf/10.1145/1219092.1219096?casa\_token=jnPh7x56VyUAAAAA:c9qNqq8XIpNvjxiVHL4n41bmh-HimPh16Tm80C2bCKcetw3tTYJs0STFM1wHjvddjb7FVtRhQqAQ}{The
  measurement calculus},'' \emph{Journal of the ACM (JACM)}, vol.~54, no.~2,
  pp. 8--es, 2007.

\bibitem{pius2010automatic}
E.~Pius,
  ``\href{https://static.epcc.ed.ac.uk/dissertations/hpc-msc/2009-2010/Einar\%20Pius.pdf}{Automatic
  parallelisation of quantum circuits using the measurement based quantum
  computing model},'' in \emph{High Performance Computing}, 2010.

\end{thebibliography}

\end{document}